\documentclass[article]{JHEP3} 

\usepackage{epsfig,multicol,bbm}
\usepackage{amsmath}
\newcommand\fverb{\setbox\pippobox=\hbox\bgroup\verb}
\newcommand\fverbdo{\egroup\medskip\noindent%
                      \fbox{\unhbox\pippobox}\ }
\newcommand{\feyn}[1]{{#1} \! \! \! /}
\newcommand{\ep}{\varepsilon}
\newcommand\fverbit{\egroup\item[\fbox{\unhbox\pippobox}]}
\newbox\pippobox

\title{Automatic generation of quarkonium amplitudes in NRQCD}

\author{Pierre Artoisenet, Fabio Maltoni \\
Centre for Particle Physics and Phenomenology (CP3) \\
Universit\'{e} Catholique de Louvain\\
Chemin du Cyclotron 2, B-1348 Louvain-la-Neuve, Belgium\\
E-mails: \email{pierre.artoisenet@uclouvain.be, fabio.maltoni@uclouvain.be}
}

\author{Tim Stelzer\\
Department of Physics, University of Illinois at Urbana-Champaign, \\ 
1110 West Green Street, Urbana, IL\ \ 61801 \\
E-mail: \email{tstelzer@uiuc.edu}
}

\preprint{CP3-07-31}            

\abstract{
We present a simple method to automatically evaluate arbitrary
tree-level amplitudes involving the production or decay of a heavy
quark pair $Q\bar Q$ in a generic $^{2S+1}L_J^{[1,8]}$ state, i.e., the
short distance coefficients appearing in the NRQCD factorization
formalism. Our approach is based on extracting the relevant
contributions from the open heavy quark-antiquark amplitudes 
through an\ expansion with respect to the quark-antiquark relative momentum 
and the application of suitable color and spin projectors. 
To illustrate the capabilities of the method and its implementation in
MadGraph a few applications to quarkonium collider 
phenomenology are presented.}

\begin{document}

\section{Introduction} 

Bound states of heavy quarks, such as $J/\psi$, $\Upsilon$ or $B_c$,
are unique systems where our understanding of QCD both at the
perturbative and non-perturbative level can be tested. Since the mid
seventies, when $J/\psi$ was discovered and asymptotic freedom invoked
to explain its narrow width, an impressive amount of experimental data
have provided continuous challenges to the theoretical community. To
date several open questions remain to be answered that concern both
production and decay mechanisms~\cite{Brambilla:2004wf}.

One of the outstanding issues is to what extent a non-relativistic
treatment of a quarkonium system is a useful starting point to
describe the production and decay of bound states of heavy quarks.
This is best tested in the case of charmonium, where the mass is not too far from $\Lambda_{QCD}$.
The appeal of such an approach, is the
possibility of embedding it in a rigorous framework, non-relativistic
QCD (NRQCD)~\cite{Bodwin:1994jh}, that allows one to make consistent
theoretical predictions which can systematically be improved. NRQCD is an
effective field theory that can be matched to full QCD.
Effects of
order $Q^2 \ge m_Q^2$  are encoded in short distance coefficients that can
be estimated by using perturbation theory.  Low $Q^2 < m_Q^2$ effects,
typically those involving hadronization, are factorized into
long-distance matrix elements, which are universal, can be organized
hierarchically in powers of $v$ and can be either measured on lattice simulations
or extracted from experimental data.

Despite its theoretical appeal and undeniable successes,
not all of the predictions of the NRQCD factorization approach have been
firmly established. Consider for example the 
color-octet contributions in the $J/\psi$ production: they seem to play a
dominant role at the Tevatron and in $\gamma \gamma$ collisions, but
they look marginal at $e^+e^-$ at low energy, in photoproduction at
HERA~\cite{Kramer:2001hh} and in fixed-target
experiments~\cite{maltoni:2006yp}. Another example is the
of $J/\psi$ 's at high-$p_T$ at the Tevatron. NRQCD predicts 
a sizeable transverse polarization 
in contrast to the latest data that now clearly indicate that
$J/\psi$'s are not transversely polarized~\cite{Abulencia:2007us}.

Given such a puzzling scenario, it is mandatory to re-examine in
details the key observables and the corresponding available theoretical 
predictions and try to systematically improve on them. Two main directions
can be followed.

The first is to calculate predictions for cross sections and relevant 
distributions at the next-to-leading order (NLO) accuracy in the strong 
coupling expansion. Even for a small number of final state partons
these calculations are technically very challenging and only a few
are currently available.  For example, there are only three
calculations relevant for describing the $p_T$ of the quarkonium
state at NLO at colliders, i.e., those whose born process is a $2\to 2$: 
the $\gamma \gamma \to ~^3S_1^{[1]} +\gamma+ X$~\cite{Klasen:2004az}, the
pioneering $\gamma p \to ~^3S_1^{[1]}+X$~\cite{Kramer:1995nb} and the recently
evaluated $pp \to ~^3S_1^{[1]}+X$~\cite{Campbell:2007ws}. In $e^+e^-$ collisions
the cross section to produce $ ~^3S_1^{[1]} c \bar c$ is the only 
inclusive charmonium cross section known at NLO~\cite{Zhang:2006ay}.

The second way is to extend the set of the typical observables
analyzed by studying new processes and also predicting
more exclusive observables. To this aim, we propose to use a
multi-purpose matrix element based generator.  This kind of codes,
which have become widely available recently, have dramatically 
reduced the burden of performing phenomonological studies at colliders. 
They automatically
create the matrix elements corresponding to a given process and then
generate unweighted parton-level events that can be passed to
standard Monte Carlo programs such as Pythia~\cite{Sjostrand:2006za} or Herwig~\cite{Corcella:2002jc}, for showering
and hadronization, and eventually to detector simulation. The
advantages of such an approach are numerous.  The most important one
is that any theoretical description (model) can be directly used in
the experimental analyses and any kind of realistic observable can
then be studied both at the theoretical and experimental (detector)
level, within the same framework. Second, since the computation of the
matrix element is automatic, virtually any process can be simulated at
tree level accuracy, opening up the possibility of straightforward
experimental analyses of potentially interesting ideas.

The purpose of this work is to present the first step in this 
direction, i.e., an algorithm for the automatic matrix element
generation of amplitudes involving the production (or the decay) of a
short-distance quarkonium state with arbitrary spin, angular momentum,
and color quantum numbers. The algorithm has been implemented in the
matrix element generator MadGraph~\cite{Stelzer:1994ta}. 
Inclusion of the matrix elements into the automatic phase space 
integrator and event generator MadEvent~\cite{Maltoni:2002qb,Alwall:2007st} 
is ongoing and will be presented in a second
publication.  The plan of the paper is as follows. In
Section~\ref{one} we describe the approach followed to project open
quark-antiquark amplitudes into those for a given $^{2S+1}L_J^{[1,8]}$
state and the validation performed.  In Section~\ref{sec:app}
we present three applications to collider phenomenology. We draw our
conclusions 
in the final section.

\section{Quarkonium amplitudes in NRQCD}
\label{one}
In the framework of Non Relativistic QCD, the cross
section associated to a heavy quarkonium production process can be
factorized as
\begin{equation}
\sigma(ij \rightarrow \mathcal{Q}+X) = \sum_{n}  \hat \sigma (ij \rightarrow Q \bar{Q}(n)+X ) \langle \mathcal{O}^{\mathcal{Q}} (n) \rangle\,,
\label{NRQCDcrosssection}
\end{equation}
where
\begin{itemize}
\item $\hat \sigma (ij \rightarrow Q \bar{Q}(n)+X )$ is the short
       distance cross section,
\item $\langle \mathcal{O}^{\mathcal{Q}} (n) \rangle$ is the long
      distance matrix element.
\end{itemize}
The short distance cross section $\hat \sigma (ij \rightarrow Q
\bar{Q}(n)+X ) $ is related to the creation of a heavy quark pair in a
quantum state $n$ and it can be computed within perturbation theory,
by expanding the amplitude in powers of $\alpha_S$ and $v$, i.e.,  the strong
coupling constant and the relative velocity $v$.  
The sum over $n$ in Eq.~(\ref{NRQCDcrosssection})
 \textit{a priori} runs over all possible non-pertubative transitions $Q \bar Q(n) \rightarrow \mathcal
Q$ and goes therefore beyond the color-singlet approximation, where
only the perturbative state with the same quantum numbers as the physical state is included. In particular, the intermediate $Q \bar Q$ state 
can be in a color-octet, the color being neutralized 
at later times.

In practice, any quarkonium squared amplitude is obtained by means of
an expansion in powers of the relative velocity $v$ between the heavy quarks. 
At leading order in $v$, the intermediate $Q \bar Q$ state can be 
specified by the spectroscopic notation
\begin{equation}
n = ~^{2S+1}L_J^{[c]}\,,
\end{equation}
where 
$S$ identifies the spin state of the heavy quark pair, 
$L$ the orbital momentum state, 
$J$ the total angular momentum state, 
and $c=1,8$ the color state.

To evaluate the short distance coefficients we use the general
algorithms for Feynman diagrams based computations available in
MadGraph~\cite{Stelzer:1994ta}. MadGraph can provide partonic helicity
amplitudes for any process in the Standard Model and beyond at the tree
level. In the case of quarkonium, we start from 
the helicity amplitude information for the open
quark-antiquark production and then apply several projection
subroutines that select the specific quantum state
numbers, $^{2S+1}L_J^{[c]}$. The projectors are universal and depend
only on the quantum numbers to be selected and not on the specific
process requested. In fact, only the color projection has to be performed
within MadGraph itself since it changes the way the amplitudes are calculated
(and in particular selects which diagrams contribute) while spin projections and 
velocity expansions are performed at runtime. 
Our method entails  that any squared matrix amplitude $|\mathcal M (ij \rightarrow Q \bar Q ( ^{2S+1}L_J^{[c]})+X)|^2$ can be generated in any of the models available in MadGraph. 
Limitations are typical to Feynman-diagram based matrix
element generators and are connected to the factorial growth in the
number of diagrams.  Let us  stress a few important points.
First, our implementation also allows the calculation of  the relativistic corrections
for S-wave state production.  Second, the employed projectors are general
enough to handle the case where the quark and the anti-quark are
of different flavor, and therefore have a different mass.  In our
approach spin correlation between the quarkonium angular momentum and
its decay products are fully  included and therefore the information
on the quarkonium polarization is kept. In Section
\ref{sec:polarisation}, we discuss two methods to disantangle the
longitudinal and transverse parts of the cross section for the
transition $Q \bar Q(^3S_1^{[1]}) \rightarrow \mathcal
Q(J^{PC}=1^{--})$.

\subsection{Projection method and its implementation}

In this section, we discuss how to combine the open-quark helicity amplitudes generated by MadGraph in order to select a specific configuration $^{2S+1}L_J^{[c]}$ for the heavy-quark pair. Our method can be seen as a generalization and automatization of that proposed in Refs.~\cite{Hagiwara:2004pf,Artoisenet:2007xi}.

First we address the computation of the color factors.  In MadGraph,
the color structure of the parton-level amplitude is obtained by
organizing the amplitude into gauge invariant subsets, corresponding 
to different color flows in the large-$N_C$ limit~\cite{Maltoni:2002mq}. 
The total amplitude is then expressed as a sum over the color flows
\begin{equation}
\mathcal M  = \sum_{m} f_m(a,b,\dots)  A_m \, ,
\end{equation} 
where the $f_m$'s are color factors (orthogonal in the large $N_c$ limit) that depend on color indices $a,b,\dots$ carried by gluons and quarks, and the $A_m$'s are gauge invariant quantities, the so-called  partial amplitudes.
These amplitudes are computed in MadGraph with the help of HELAS subroutines \cite{Murayama:1992gi}. The color factors are evaluated at the level of the squared amplitude, each color index being contracted:
\begin{equation}
|\mathcal M|^2 =\left( A_1^*,A_2^*,\dots \right) \left( \begin{array}{ccc} f_1^*f_1 & f_1^* f_2 & \dots \\ f_2^*f_1 & f_2^* f_2 & \dots \\ \vdots & \vdots & \end{array}    \right)  \left( \begin{array}{c} A_1 \\ A_2 \\ \vdots \end{array}  \right) \, .
\end{equation}
In MadGraph, the decomposition of the partonic amplitude   into partial gauge-invariant amplitudes, as well as the evaluation of the matrix of color factors,  are computed automatically for any process at tree-level.

In order to compute the color structure related to the process $ij
\rightarrow Q \bar Q ( ^{2S+1}L_J^{[c]})+X$, the quark and anti-quark  are
required to be in a given color state $[c]$.  Given that 
$ 3 \otimes \bar 3 = 1 \oplus 8$ the pair can be either in a singlet or octet state that can be obtained by using the projectors~\cite{Petrelli:1997ge}
\begin{equation}
\mathcal{C}_1=\frac{\delta_{ij}}{\sqrt{N_C}}, \quad \mathcal{C}_8=\sqrt{2} (T^a)_{ij} ,
\end{equation}
where $i$ and $j$ are the color indices of the heavy quarks.
These projectors have been implemented into MadGraph. The color structure for the quarkonium production amplitude is then computed using the same technique as for the open quark production. \\

Next we consider the projection of the amplitude onto a definite total
spin state of the heavy quark pair.  The spin projectors were first derived
in Ref. \cite{Berger:1980ni}, and we employ the
normalization of Ref.~\cite{Petrelli:1997ge}.  Since the code is also
designed for the production of heavy quarkonia of mixed flavours, we
distinguish the mass of the quark ($m_1$) from  that of the
anti-quark ($m_2$).  $p_1$ ($p_2$),  the momentum of the heavy quark
 (anti-quark)  can be expressed in
 terms of the total momentum $P$ and the relative momentum $p$:
\begin{equation}
p_1=\frac{m_1}{m_1+m_2} P+p, \quad p_2=\frac{m_2}{m_1+m_2}P -p \, .
\label{kin_definition}
\end{equation}
The partial helicity amplitudes can be combined to yield a specific spin state for the heavy quark pair:
\begin{equation}
\mathcal M(ij \to Q \bar Q (S,\lambda)+X)= \sum_{\lambda_1,\lambda_2} N(\lambda|
\lambda_1,\lambda_2)
\mathcal M(ij \to Q(\lambda_1) \bar
Q(\lambda_2) + X)\,,
\label{spinproj1}
\end{equation}
where  the Clebsch-Gordant coefficients $N(\lambda|\lambda_1,\lambda_2)$ can 
be written in terms of the   heavy quark spinors~\footnote{We use the normalization $\bar{u}(p_i,\lambda_i)u(p_i,\lambda_i)=2m_i=- \bar v(p_i,\lambda_i) v(p_i,\lambda_i)$.},
\begin{equation}  
N(\lambda| \lambda_1,\lambda_2) =\frac{1}{\sqrt{8m_1m_2}}
 \bar{v} (p_2,\lambda_2)
  \Gamma_S u (p_1,\lambda_1) \,\, ,
\label{CGcoeff1}
\end{equation}
with $\Gamma_{S=1} = \epsilon^\lambda_{\mu} \gamma^\mu $ for a spin-one $Q \bar Q$ state ($\epsilon^\lambda_{\mu}$ being the polarization vector), and $\Gamma_{S=0} =  \gamma_5 $  for a spin-zero $Q \bar Q$ state.
Note that this expression is accurate only at leading order in the relative momentum  $p$. The helicity amplitudes $\mathcal M(ij \to Q(\lambda_1) \bar
Q(\lambda_2) + X)$ are computed by  MadGraph in a specific Dirac representation employed in the HELAS subroutines~\cite{Murayama:1992gi}. These subroutines can also be invoked for the evaluation of the current in the rhs of Eq.~(\ref{CGcoeff1}). 
The spin projection formula in Eq. (\ref{spinproj1}) can then be easily 
incorporated in a  generic numerical algorithm. \\

We then consider the projection onto a specific orbital state.  For
$S$-wave state production at leading order in $v$, one can simply set
 $p=0$ in the short distance amplitude. For $P$-wave state, the
 leading order contribution in $v$ is given by the derivative of the
 amplitude with respect to the relative momentum. We approach this
 derivative numerically by the quotient
\begin{equation}
\frac{\mathcal M(\Delta p^i)-\mathcal M(0)}{\Delta p^i}
\end{equation}
in the quarkonium rest frame\footnote{Wherever we consider a non-zero relative momentum, we replace $m_i$ by $E_i$ (the energy of the quark $i$) in Eq.~(\ref{kin_definition}) in order to enforce the mass-shell condition for the heavy quarks.}.

Eventually, for spin-one $P$-wave states, the spin index is combined with the orbital index in order to select a given angular momentum state, $J=0,1$ or $2$. 

\subsection{Relativistic correction to $S$-wave state production}
\label{sec:relcorr}

To evaluate the relativistic corrections to $S$-wave state
production, the relative momentum $p$ must be kept different from zero in
the amplitude. In this case, we need an expression of the spin projectors
accurate to all orders in $p$. According to the
result of Ref.~\cite{Bodwin:2002hg}, the Clebsch-Gordan coefficients can be written as
\begin{equation}  
N(\lambda| \lambda_1,\lambda_2) =\frac{1}{\sqrt{2} (E+m)}
 \bar{v} (p_2,\lambda_2)
 \frac{\feyn{P}+2E}{4E} \Gamma_S u (p_1,\lambda_1) \,\, ,
\label{CGcoeff2}
\end{equation}
where $E=\frac{\sqrt{P^2}}{2}$. In this case we assume that the quark and the anti-quark are of the same flavour ($m_1=m_2=m$). The contribution to an $S$-wave
configuration is selected by projecting the amplitude onto the
spherical harmonic $Y^{m=0}_{l=0}$. This amounts to averaging the
amplitude over the direction of the relative momentum.
The short distance amplitude is then expanded in powers of 
$\frac{\boldsymbol p^2}{m^2}$, each term in the expansion being associated with an
independent non-perturbative long distance matrix element.

It has been shown recently that the relativistic corrections arising from the wave function for $S$-wave state can be resummed to all orders in $v$~\cite{Bodwin:2006ke,Bodwin:2006dn}. Thanks  to the following relation among the long distance matrix elements
\begin{equation}
\frac{\langle 0 |\chi^\dagger \kappa (-\nabla^2)^n \psi |\mathcal Q  \rangle}{\langle 0 |\chi^\dagger \kappa \psi |\mathcal Q  \rangle}=\left (\frac{\langle 0 |\chi^\dagger \kappa(-\nabla^2) \psi |\mathcal Q  \rangle}{\langle 0 |\chi^\dagger \kappa \psi |\mathcal Q  \rangle} \right)^n \, ,
\label{cross_sec_with_rel_corr}
\end{equation}
where $\kappa$ is either $1$ for spin-zero $Q \bar Q$ states or the Pauli matrix $\boldsymbol{ \sigma}$ for spin-one $Q \bar Q$ states,
 the cross section can be expressed in term of one long distance matrix element:
\begin{equation}
\sigma = \hat \sigma (<\boldsymbol q^2>) \langle 0 |\chi^\dagger \kappa \psi |\mathcal Q  \rangle \langle \mathcal Q |\psi^\dagger \kappa \chi |0 \rangle\,,
\label{cross_sec_with_rel_corr2}
\end{equation}
where the short distance coefficient has been evaluated at the relative momentum
\begin{equation}
<\boldsymbol q^2>=\frac{\langle 0 |\chi^\dagger \kappa (-\nabla^2) \psi |\mathcal Q  \rangle}{\langle 0 |\chi^\dagger \kappa \psi |\mathcal Q  \rangle} \, .
\end{equation}
The inclusion of the relativistic corrections arising from the wave function has been implemented in the code, using Eq. (\ref{cross_sec_with_rel_corr2}).

\subsection{Polarization of vector-like quarkonium states}
\label{sec:polarisation}

The polarization of  vector-like quarkonium states is an interesting exclusive
quantity that can provide further information on the production mechanism since it 
can be measured accurately at colliders. On the theoretical side, 
the transverse and longitudinal parts of the cross section can be disentangled
by using the explicit basis of polarization vectors $(\ep_L, \ep_{T1}, \ep_{T2})$, defined by the equations
\begin{equation}
\ep_i . \ep_j = - \delta_{ij}, \quad \ep_i . P = 0, \quad \boldsymbol{\ep}_L. \boldsymbol{P}=0 .
\label{polarisationbasis}
\end{equation}
This representation is obviously Lorentz-frame dependent. 
Usually, the three-vector $\boldsymbol{P}$ is evaluated in the lab frame
in Eq.~(\ref{polarisationbasis}) but other useful frames exist \cite{Beneke:1998re}.

At hadron colliders a $^3S_1$ quarkonium state is observed through
its decay into leptons. 
The polarization of the quarkonium can indeed be determined
by analyzing the angular distribution of the leptons.
Defining $\theta$ as the angle between the $\ell^+$
direction in the quarkonium rest frame and the quarkonium direction in
the laboratory frame, the normalized angular distribution $I(\cos \theta)$ is
\begin{equation} 
\label{angulardist} 
I(\cos \theta) =
\frac{3}{2(\alpha+3)} (1+\alpha \, \cos^2 \theta)\,, 
\end{equation}
 where
the relation between $\alpha$ and the polarization state of the
quarkonium is 
\begin{equation}
 \label{defalpha}
\alpha=\frac{\sigma_T-2\sigma_L}{\sigma_T+2\sigma_L} \,. 
\end{equation}

In fact, a more general way to correctly account for the polarization information
is to decay the vector-like quarkonium state into leptons at the matrix element level.
This can be achieved by replacing the polarization vector of
the quarkonium by the leptonic current
\begin{equation}
\frac{\sqrt{3}}{8m \sqrt{\pi}}\bar{u}_{\ell^-}(k_1,\lambda_1)
\gamma_\mu v_{\ell^+}(k_2,\lambda_2) \,.
\end{equation}
and has been set up as an option in our code. Once the vector-like
quarkonium decay is required by the user,  events are generated that
include the information on the lepton momenta too. This method has
two main advantages. First it is directly connected to the
experimental analysis. The acceptance of the detector may affect the relation between the
measured value of the parameter $\alpha$ and the perturbative polarized
cross sections $\sigma_T$, $\sigma_L$. However, even when cuts are
applied, the angular distribution of the leptons is
the physical observable that can be used to compare theory and
measurements. Second, it is Lorentz invariant and therefore does not
impose an a priori choice for the frame where the polarization information
is extracted.

\subsection{Validation and examples}

\subsubsection{Preliminary checks}

In order to gain confidence in our implementation, we have performed 
several checks listed below.
\begin{itemize}
\item Gauge invariance has been systematically verified for all the processes considered
by using longitudinal polarization for the external photons or gluons.
\item Numerical cancellation among diagrams has been checked for amplitudes
that vanish by symmetry considerations:
\begin{eqnarray}
&&A( ^1S_0^{[1]} + (2k+1)\,{\rm photons}) = 0, \nonumber \\
&&A( ^3S_1^{[1]} + 2k \,{\rm photons}) = 0, \nonumber \\ 
&&A( ^1P_1^{[1]} + 2k\,{\rm photons}) = 0, \nonumber \\
&&A( ^3P_1^{[1]} + 2k \,{\rm photons}) = 0, \nonumber \\ 
&&A( ^3P_{0,2}^{[1]} + (2k+1)\,{\rm photons}) = 0,\nonumber 
\end{eqnarray}   
with $k=1,2,3$.

\item We have compared our numerical 
amplitudes against the analytic results point-by-point in phase
space for ($i,j,k=$ quarks or gluons):
\begin{enumerate} 
\item $i j \to {\cal Q}  k$   
for all $S$- and $P$-wave states, both color-singlet and color-octet amplitudes, 
~\cite{Berger:1980ni}\cite{Cho:1995ce}\cite{Cho:1995vh};
\item $i j \to {\cal Q} V$, with $V=Z,W$ for the relevant 
$S$- and $P$-wave states, both color-singlet and color-octet 
amplitudes~\cite{Kniehl:2002wd};
\item $i j \to {\cal Q} \phi$,  $\phi$ being a scalar or pseudo-scalar 
for the relevant $S$- and $P$-wave states, both color-singlet and 
color-octet amplitudes \cite{Kniehl:2004fa};
\end{enumerate}
The agreement for $S$ wave amplitudes is at the machine precision, 
while for $P$-waves, which are obtained through a numerical derivative, 
it is typically at the $10^{-5}$ level.

\end{itemize}

\subsubsection{Example 1: $B_c$ production at various colliders}

\begin{table}

\[
\begin{array}{|c|cccccc}

 (fb) & ^1S_0^{[1]} & ^3S_1^{[1]} & ^1P_1^{[1]} & ^3P_0^{[1]} & ^3P_1^{[1]} & ^3P_2^{[1]} \\
\hline

e^+e^- @m_Z
& 1.69 \cdot 10^{3}  & 2.37 \cdot  10^{3}  & 1.78 \cdot 10^{2} &1.08 \cdot 10^{2} &  2.23 \cdot 10^{2} & 2.40 \cdot 10^{2} \\

\gamma \gamma @ \textrm{LEP II}
&0.519 &5.23 & 0.162 &2.69 \cdot 10^{-2} &5.80 \cdot 10^{-2} &0.266  \\

\gamma p @ \textrm{HERA}
& 3.66 \cdot 10^2 & 1.76 \cdot 10^{3} & 85.4  & 21.7 & 51.7 & 2.02 \cdot 10^2  \\

\begin{array}{cc}
pp @ \textrm{LHC} &
\begin{array}{c}
g  g \\ q \bar q
\end{array}
\end{array}
&
\begin{array}{c}   3.94 \cdot 10^{7}  \\   1.37 \cdot 10^{5}     \end{array}
&
\begin{array}{c}   9.83 \cdot 10^{7}  \\   8.34 \cdot 10^{5}     \end{array}
&
\begin{array}{c}    5.20 \cdot 10^{6} \\   2.95 \cdot  10^{4}     \end{array}
&
\begin{array}{c}    1.82 \cdot 10^{6} \\   1.10 \cdot  10^4     \end{array}
&
\begin{array}{c}    4.40 \cdot 10^{6} \\   2.69 \cdot 10^4      \end{array}
&
\begin{array}{c}   1.05 \cdot 10^{7}  \\   7.26 \cdot 10^4     \end{array}
\\

\begin{array}{cc}
p\bar p @ \textrm{Tev II} &
\begin{array}{c}
g  g \\ q \bar q
\end{array}

\end{array}
&
\begin{array}{c} 2.56 \cdot 10^{6}    \\ 2.64 \cdot 10^4    \end{array}
&
\begin{array}{c}  6.30 \cdot 10^{6}    \\ 1.62 \cdot 10^5     \end{array}
&
\begin{array}{c}  3.28 \cdot 10^{5}   \\ 5.70 \cdot 10^3    \end{array}
&
\begin{array}{c}  1.24 \cdot 10^{5}   \\  2.12 \cdot 10^3   \end{array}
&
\begin{array}{c}  2.83 \cdot 10^{5}    \\  5.21 \cdot 10^3   \end{array}
&
\begin{array}{c}   6.62 \cdot 10^{5}  \\   1.41 \cdot 10^4  \end{array}

\end{array}
\]

\label{tab:bc1}
\caption{Cross sections (femtobarn) for the color-singlet contributions to 
the inclusive production of $B_c$. }

\end{table}

\begin{table}
\[
\begin{array}{|c|cccccc}

(fb) & ^1S_0^{[8]} & ^3S_1^{[8]} & ^1P_1^{[8]} & ^3P_0^{[8]} & ^3P_1^{[8]} & ^3P_2^{[8]} \\
\hline

e^+e^- @ m_Z
&1.59 & 2.22   & 0.167  & 0.102 & 0.209 & 0.225    \\
\gamma \gamma @ \textrm{LEP II}
&4.87  \cdot 10^{-4} &4.90 \cdot 10^{-3} &1.52 \cdot 10^{-4} &2.52 \cdot 10^{-5} &5.44  \cdot 10^{-5} &2.49 \cdot 10^{-4} \\

\gamma p @ \textrm{HERA}
& 1.18   & 8.46 &0.506 &7.64 \cdot 10^{-2} & 0.244 &1.61  \\

\begin{array}{cc}
pp @ \textrm{LHC} &
\begin{array}{c} g  g \\ q \bar q \end{array}
\end{array}
&
\begin{array}{c} 4.11 \cdot 10^{5} \\ 1.03  \cdot 10^{3} \end{array}
&
\begin{array}{c}  1.79 \cdot 10^{6}\\ 7.03 \cdot 10^{3} \end{array}
&
\begin{array}{c} 1.17 \cdot 10^{5} \\ 251 \end{array}
&
\begin{array}{c} 1.38 \cdot 10^{4} \\ 30.7 \end{array}
&
\begin{array}{c} 6.23 \cdot 10^{4} \\  174  \end{array}
&
\begin{array}{c} 2.29 \cdot 10^{5} \\ 616  \end{array}
\\

\begin{array}{cc}
p\bar p @ \textrm{Tev II} &
\begin{array}{c}
g  g \\ q \bar q
\end{array}
\end{array}
&
\begin{array}{c} 2.85 \cdot 10^{4} \\  199    \end{array}
&
\begin{array}{c}  1.26 \cdot 10^{5} \\  1.37 \cdot  10^3     \end{array}
&
\begin{array}{c}  8.13 \cdot 10^{3} \\  48.6    \end{array}
&
\begin{array}{c}   9.80 \cdot 10^{2} \\   5.95  \end{array}
&
\begin{array}{c}   4.25 \cdot 10^{3} \\  33.7   \end{array}
&
\begin{array}{c}   1.57 \cdot 10^{4} \\   120  \end{array}

\end{array}
\]

\label{tab:bc8}
\caption{Cross sections (femtobarn) for the color-octet contributions to 
the inclusive production of $B_c$. }

\end{table}

$B_c$ production has been extensively studied at various colliders and 
the many available results provide us with a further testing ground
for our code. 

We present the results for inclusive cross sections at several colliders
for all the $S$ and $P$, both color-singlet and color-octet, states.
Since our purpose is to provide reference numbers, we use a very simple set of
input parameters common to all processes:
\begin{itemize}
\item $m_b=4.9$  GeV, $m_c=1.5$  GeV, $m_{B_c}=6.4$  GeV 
\item $\langle O\left( ^{2S+1}S_J^{[1]} \right) \rangle =(2J+1)\, 0.736$  GeV$^3$ 
\item $\langle O\left(^{2S+1}P_J^{[1]} \right)\rangle =(2J+1) \, 0.287$  GeV$^5$ 
\item $\langle O^{[8]} \rangle =0.01 \langle O^{[1]} \rangle$
\item $\mu_{epa}=\mu_F=\mu_R=12.8$ GeV 
\item $\alpha_S(\mu_R)=0.189$
\item pdf set: cteq6l1
\item $\alpha_{EM}=\frac{1}{137}$
\end{itemize}

Our results are summarized in Tables 1 and 2.
For the processes $gg \rightarrow B_c b \bar c$ we compared 
with Refs. \cite{Chang:2003cq,Chang:2005hq}, and found agreement 
for all intermediate states. For color-singlet $S$-wave state 
production in $\gamma g$ interaction, our results agree with 
those of Ref.~\cite{Berezhnoy:1997er}.
However, some of the results in the table are new.

\subsubsection{Example 2: Higgs decay into $\Upsilon + X $.}

In this section we illustrate through a very simple example 
how a multi-purpose matrix element code allows  
to quickly assess the phenomenological relevance of new ideas. 

A light Higgs of $m_H\lesssim 140$ GeV in the Standard
Model as well as in extensions such as in SUSY or in generic 2HDM,
decays predominantly into $b\bar b$. Assuming a $2\to 1$ production
mechanism at hadron colliders, such $gg$ or $b\bar b$ fusion, a two $b$-jet
signature is obtained. Such events are extremely difficult to trigger
on and to select from the enormous QCD two-jet background. In fact, in
this mass range, the golden channel for discovery is $H\to \gamma
\gamma$ which has a branching ratio of the order of only 10$^{-3}$ in
the SM but has a very clean signature: it can be triggered on and the
invariant mass determination is very accurate, giving the possibility
of reconstructing a small BW peak over a very large background.
However, in some BSM scenarios, such as for example at large $\tan
\beta$ in SUSY or in a generic 2HDM, the branching ratio to two
photons is highly suppressed and other decay modes need to be
considered.

The presence of a $\Upsilon$
in the final state could help both in the triggering (through the $\mu^+
\mu^-$ pair) and the invariant mass reconstruction and give a viable discovery mode. Let us consider two cases.
\FIGURE[t]{
\epsfig{file=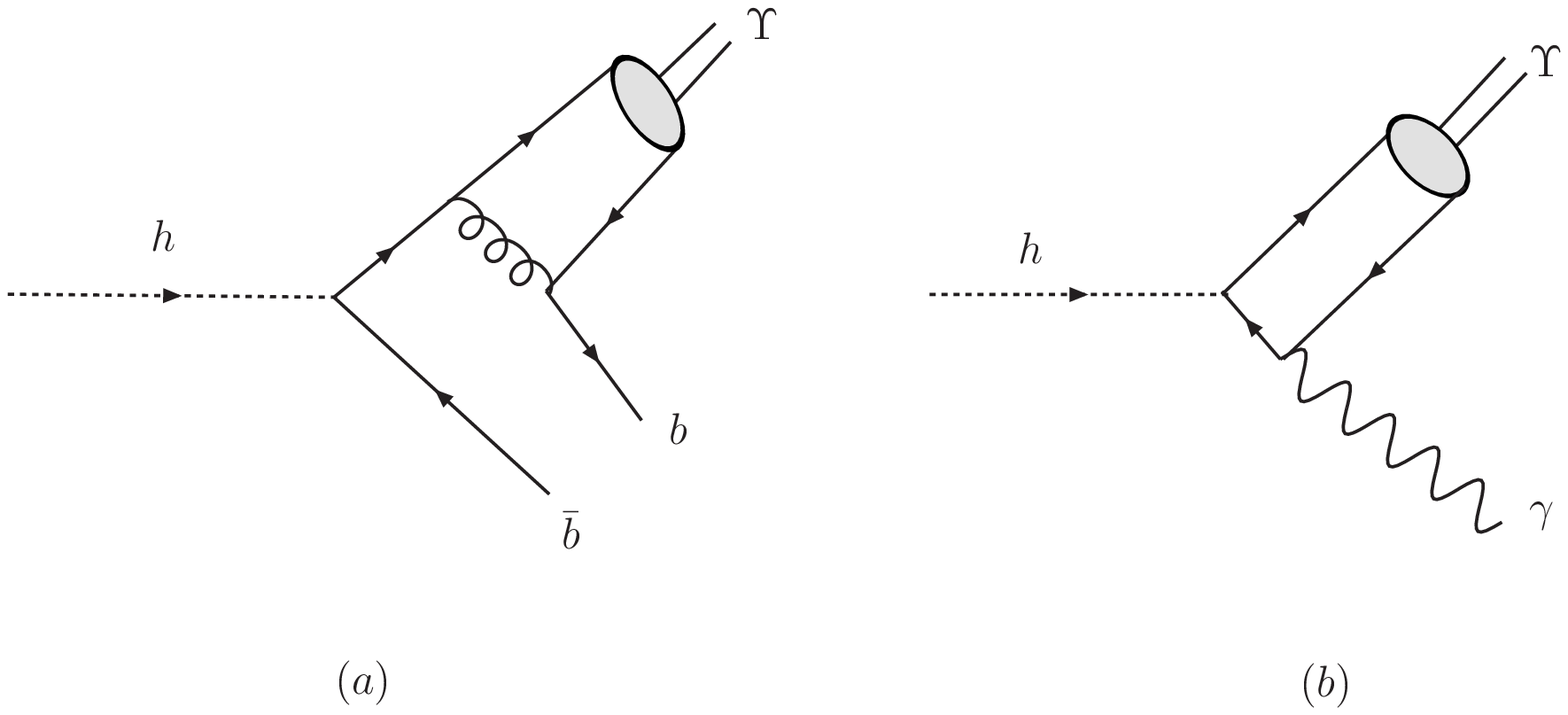   , width=10cm}
\caption{Higgs decays into $\Upsilon$: in association with a $b \bar b$  pair (a)
and with a photon (b).}
\label{fig:higgsdecay}
}
As the simplest decay mode, consider $H \to \Upsilon b \bar b$, 
Fig~\ref{fig:higgsdecay} (left).  This is 
analogous to $Z\to \Upsilon b\bar b$, which can also be
described by the fragmentation function approach. The exact
matrix element calculation gives: 
\begin{eqnarray}
\frac{\Gamma( H \to \Upsilon b \bar b)}
{\Gamma( H\to  b \bar b)}& = & 1.56 \cdot  10^{-5} \,,
\end{eqnarray}
where we used $m_H=120$ GeV, $\langle O_\Upsilon(^3S_1^{[1]}) \rangle = 9.28$ GeV$^3$ and $\alpha_S=0.118$.

Another, potentially cleaner mode, which could allow for an excellent invariant
mass resolution, is the decay $H\to \gamma \Upsilon$, Fig~\ref{fig:higgsdecay} (right) 
Note that the crossed  decay  $ \Upsilon \to \gamma H$ has been considered before in the
literature~\cite{McElrath:2005bp,  Mangano:2007gi,McElrath:2007sa,SanchisLozano:2007wv} as a search mode for a light scalar or pseudoscalar and it is 
known at the NLO accuracy~\cite{Nason:1986}. A straightforward calculation gives:
\begin{eqnarray}
\frac{\Gamma(H \to \Upsilon \gamma)}
{\Gamma(H \to b \bar b)}& = & 6.17 \cdot 10^{-7} \,,
\end{eqnarray}
where we used $m_H=120$ GeV, $\langle O_\Upsilon(^3S_1^{[1]}) \rangle = 9.28$ GeV$^3$ and $\alpha_{EM}=1/137$. 

The BR's for both  decay modes in the SM (and also in SUSY, since possible large
$\tan \beta$ enhacements cancel in the ratio of the widths), 
are found to be rather small, also considering that the branching ratio
of the quarkonium state into leptons has yet to be included. 

\section{Applications}
\label{sec:app}

\subsection{$\gamma \gamma \to J/\psi +X$ at LEP II}

\
\FIGURE[t]{
\epsfig{file=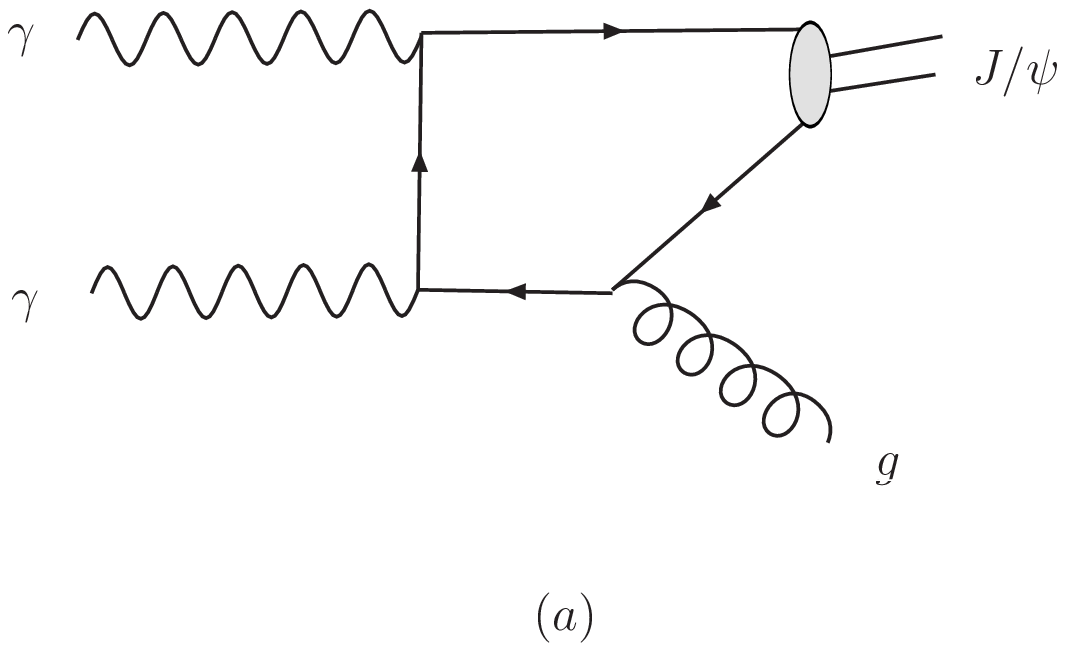 ,width=6cm}
\hspace*{.5cm}
\epsfig{file=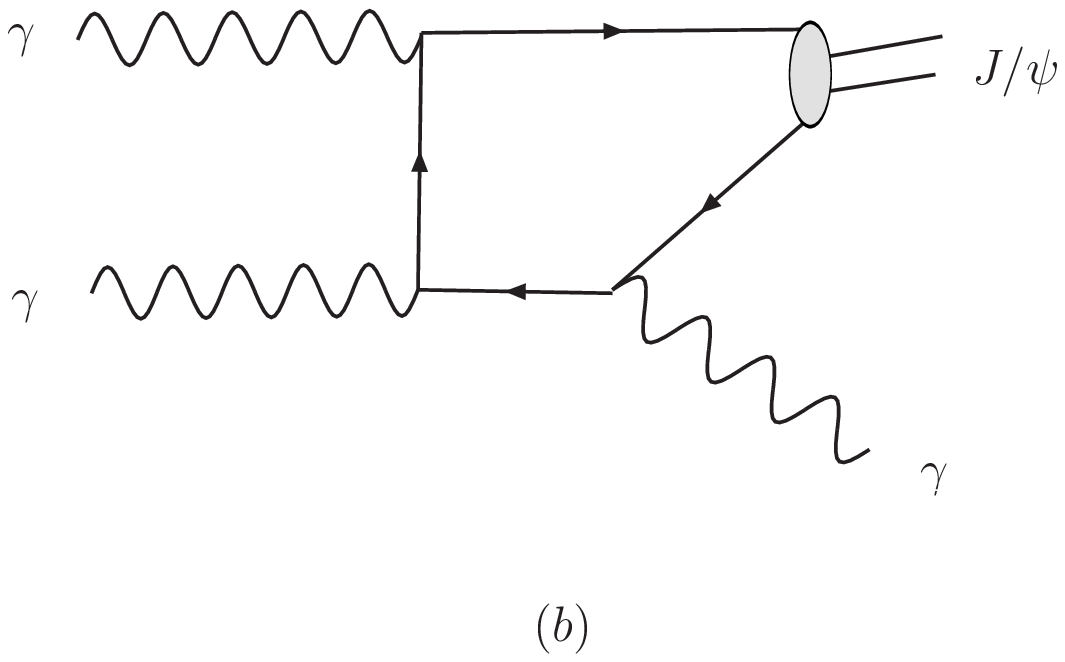   , width=6cm}\\
\epsfig{file=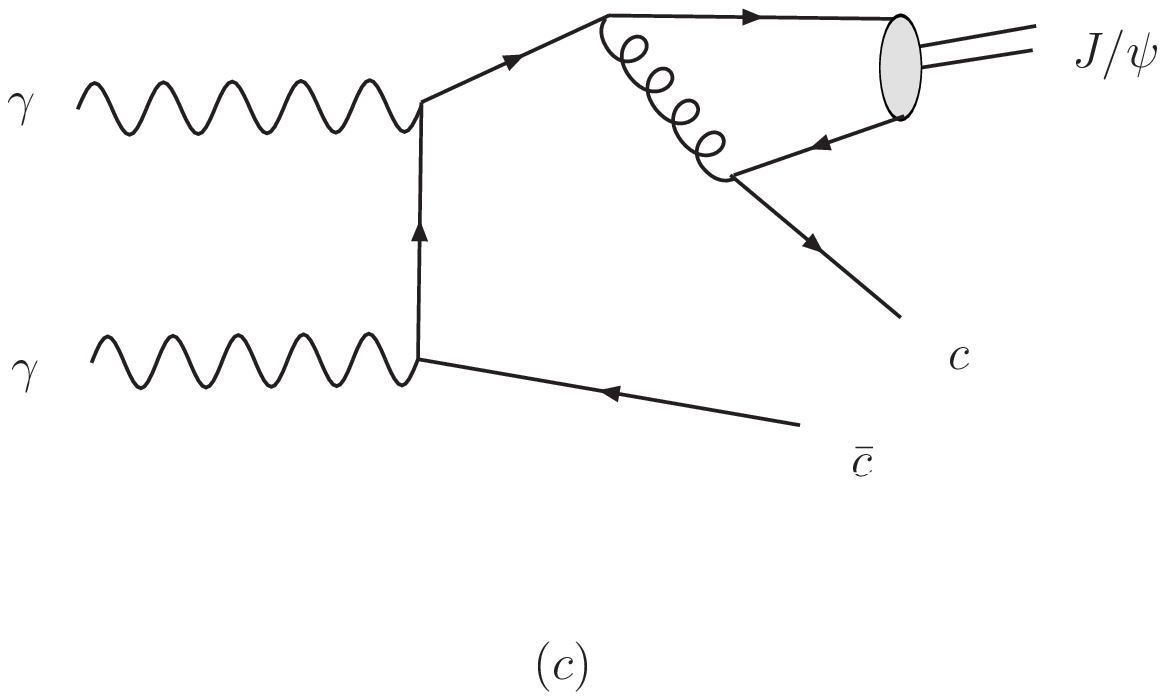   , width=6cm}
\hspace*{.5cm}
\epsfig{file=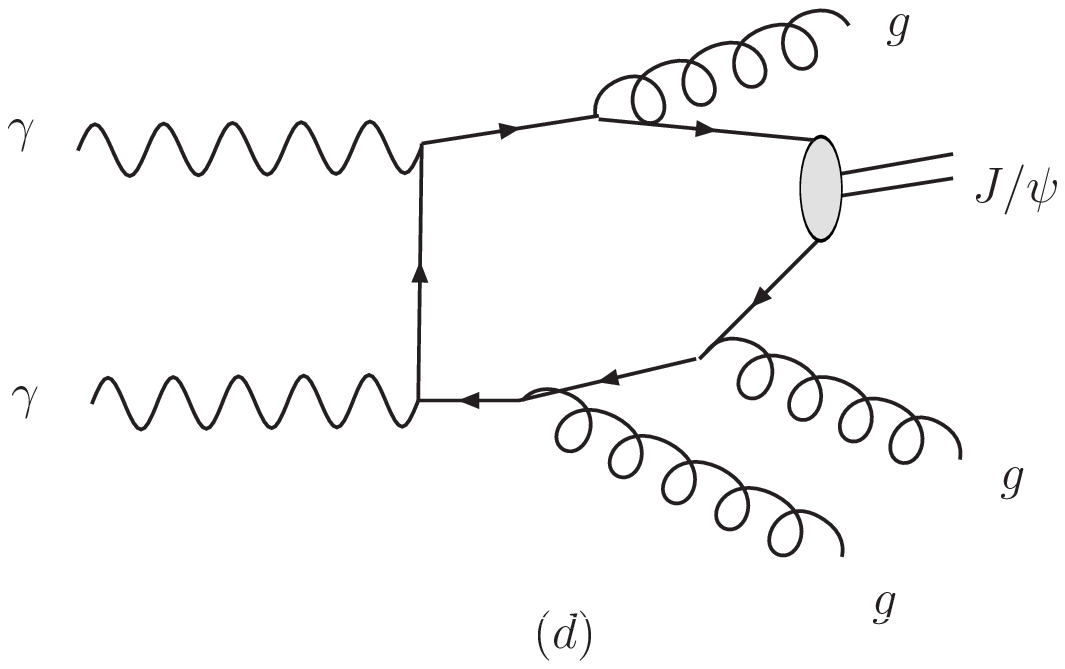   , width=6cm}
\caption{$J/\psi$ direct photoproduction. Production mechanism (a) proceeds via a $^3S_1^{[8]}$ state, while the photon associated (b), the $c\bar c$ associated and the multi-gluon (d) mechanisms can proceed via a color-singlet transition.}
\label{fig:aapsidiagr}
}

$J/\psi$ production in $\gamma \gamma$ collisions at LEP II 
($\sqrt{S}=196$ GeV) has been
studied in Ref~\cite{Klasen:2001cu}.  In that paper, the direct ($\gamma
\gamma$), the single resolved ($\gamma g$) and the double resolved ($gg$)
components have been taken into account.  However, the only direct
contribution that has been considered is the color-octet transition
\begin{equation}
\gamma \gamma \rightarrow c \bar c\left(  ^3S_1^{[8]} \right) g\,.
\label{aa_octet}
\end{equation} 
We now address the evaluation of the color-singlet $J/ \psi$ production in
(unresolved)  $\gamma \gamma$ interactions. 
Due to color conservation, the final states $c \bar c\left(  ^3S_1^{[1]} \right)g $ is forbidden.
But if we turn the gluon into a photon, the corresponding $\alpha_{EM}^3$-process is allowed. 
Charge conjugation conservation forbids the final state
$c \bar c\left(  ^3S_1^{[1]} \right)gg $. So the only color-singlet 
process at $\alpha_{EM}^2 \alpha_S^2$ is the associated production.
Finally, the channel $J/\psi +$ light partons occurs at order $\alpha_{EM}^2 \alpha_S^3$.
  So the Born-level 
contributions for the color-singlet production are given by 
\begin{equation}
\gamma \gamma \rightarrow c \bar c\left(  ^3S_1^{[1]} \right) \gamma ,\qquad
\gamma \gamma \rightarrow c \bar c\left(  ^3S_1^{[1]} \right) c \bar c ,\qquad
\gamma \gamma \rightarrow c \bar c\left(  ^3S_1^{[1]} \right) ggg. 
\label{aa_singlet}
\end{equation}
Note also that the final state $c \bar c\left(  ^3S_1^{[1]} \right) gq \bar q$
is also allowed, but is not finite: the fragmentation part $\gamma \rightarrow
q \bar q \rightarrow g$ is effectively included in the single-resolved contributions.

Using the Weisz\"acker-Williams approximation at LEP2,
$\alpha_S=0.189$, $m_{J/\psi}=2m_c=3$ GeV, $\alpha_{EM}=1/137$, $\langle
O\left(^3S_1^{[1]} \right) \rangle=1.16$ GeV$^3$, $\langle
O\left(^3S_1^{[8]} \right) \rangle=1.06 \cdot 10^{-2}$ GeV$^3$, we
obtain the $P_T$ spectra displayed in Fig. \ref{fig:aapsiX} for the
color-singlet and color-octet processes. We find that the associated
$c\bar c$ production is by far the dominant one for $p_T(J/\psi) > 5$ GeV.

\FIGURE[ht]{
\epsfig{file=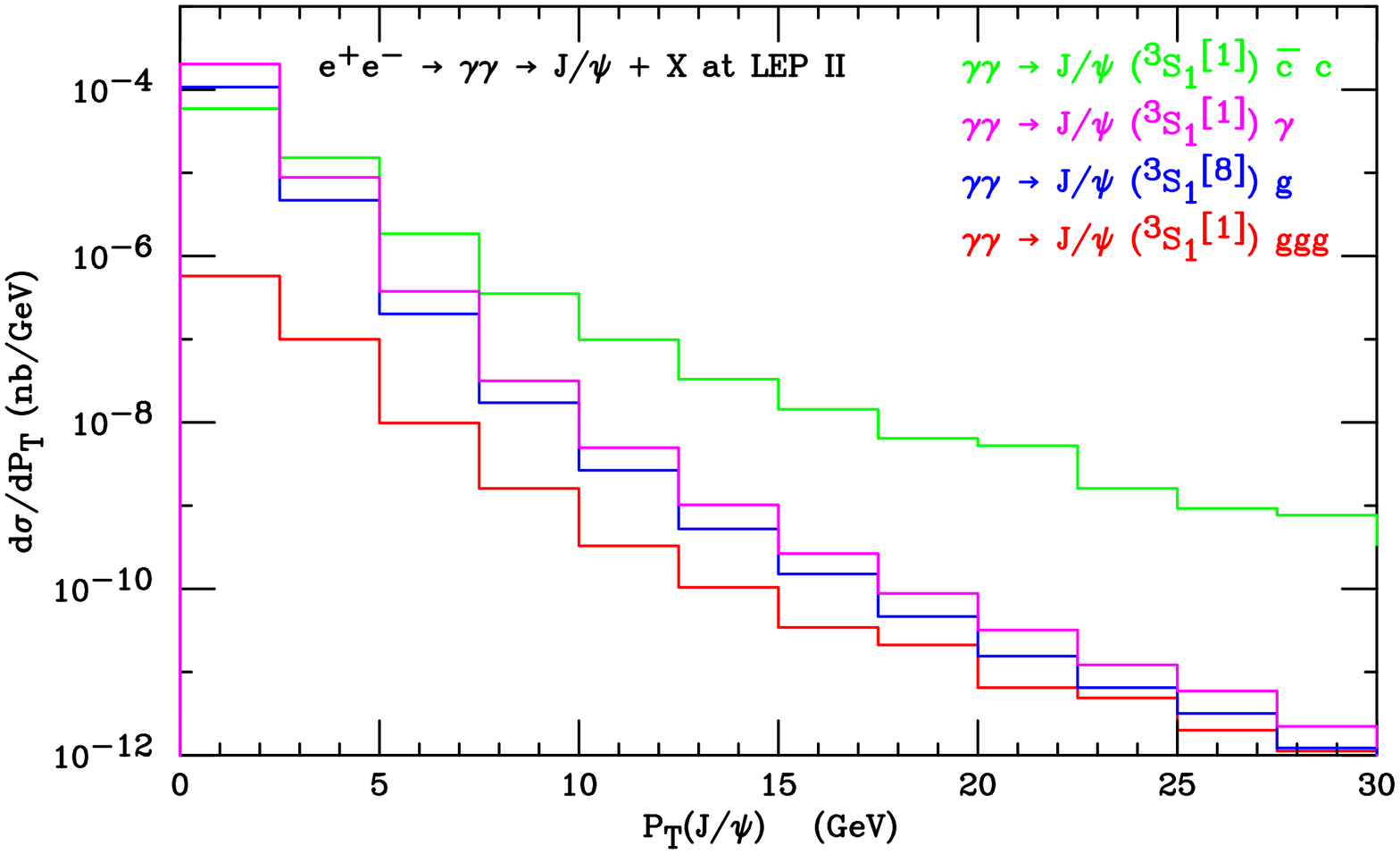,angle=90, width=12cm}
\caption{Transverse momentum distributions for $J/\psi$ production in $\gamma \gamma$ collisions}
\label{fig:aapsiX}
}

\subsection{$e^+ e^- \to \eta_c +X$ at $\sqrt{s}=10.6$ GeV}

The production of an $\eta_c$ state from $e^+e^-$ collisions has been
studied in the context of associated production~\cite{Liu:2003jj}. 
Using the inputs $\alpha_S=0.26$, $\alpha_{EM}=\frac{1}{137}$, $\langle
O\left(^1S_0^{[1]} \right) \rangle=0.387$ GeV$^3$, we obtain
\begin{equation}
\sigma (e^+ e^- \to \eta_c +c \bar c )= 58.7 \, \textrm{fb}\,,
\end{equation}
in agreement with the result obtained in Ref.~\cite{Liu:2003jj}.

 The contribution to $\eta_c$ production coming from the associated
production with  light partons has not been studied so far. 
 The reason is that for color-singlet production, neglecting
 the exchange of an off-shell $Z$,  there is
 no such processes at order $\alpha_{EM}^2\alpha_S^2$, due to color
 and charge conjugation conservation. However, at order $\alpha_{EM}^2\alpha_S^3$ 
the following processes can occur:
\begin{equation}
\label{etac_ijk}
e^+e^- \rightarrow \eta_c ggg, \qquad e^+e^- \rightarrow \eta_c q\bar q g\,,
\end{equation}
which are finite and could give a non-negligible cross section due 
to the larger phase space available.  This calculation is straightforward.
The code generates the matrix elements for the subprocesses in (\ref{etac_ijk}), 
whereas an analytical computation would have been cumbersome.  
Using a very simple phase-space generator, we checked that 
the channel $\eta_c +$ light partons constitutes approximately $10\%$ of 
the inclusive cross section.
The plot in Fig.~\ref{fig:etacX} displays the differential cross
section with respect to $z=\frac{2 |\boldsymbol{P}_{\eta_c}|}{\sqrt{s}}$, 
the fraction of momentum taken by the $\eta_c$. The total cross sections are given by
\begin{equation}
\sigma(e^+ e^- \rightarrow \eta_c ggg )=3.72 \, \textrm{fb}  , \quad \sigma(e^+ e^- \rightarrow \eta_c q \bar q g )=1.63 \, \textrm{fb}
\end{equation}
\
\FIGURE[t]{
\epsfig{file=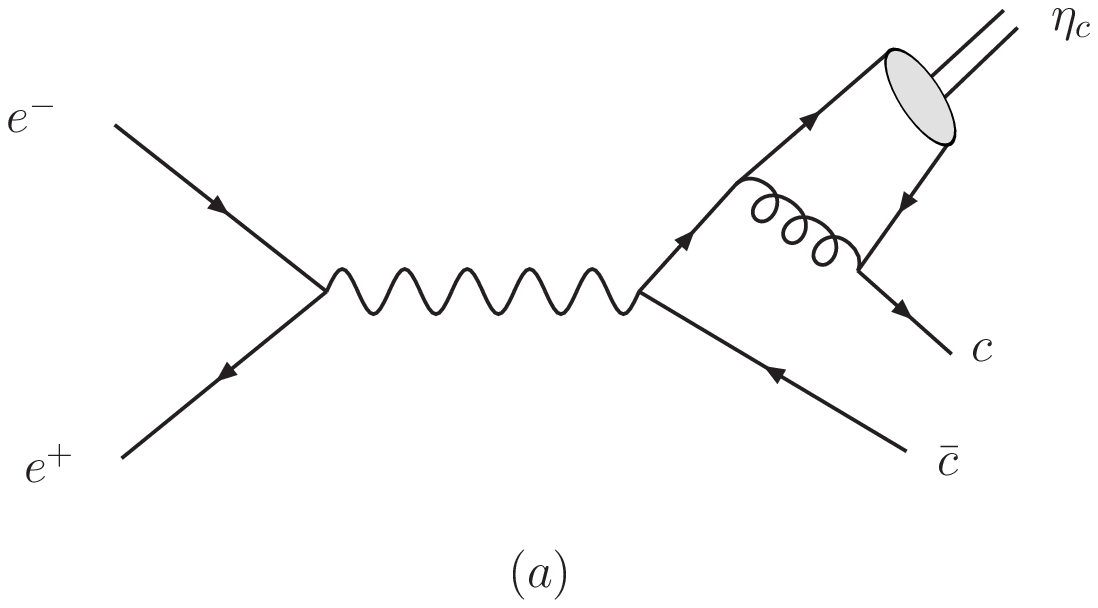 ,width=5cm}
\hspace*{.3cm}
\epsfig{file=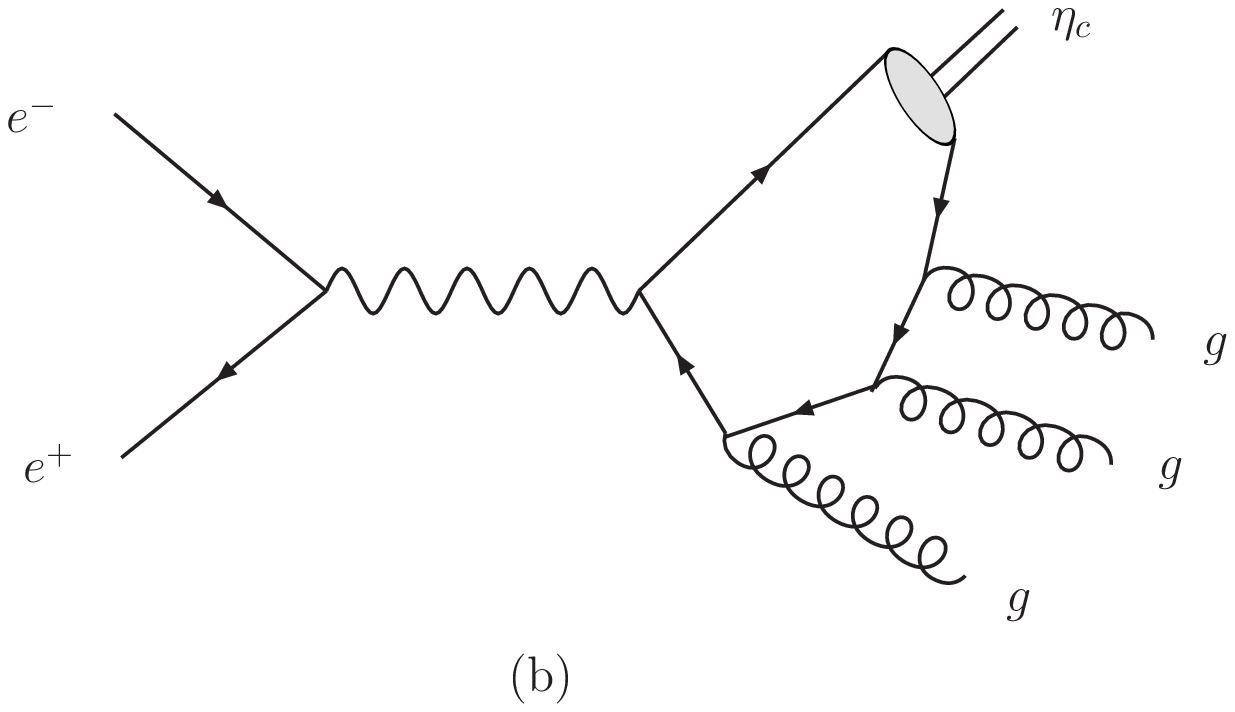   , width=5cm}
\hspace*{.3cm}
\epsfig{file=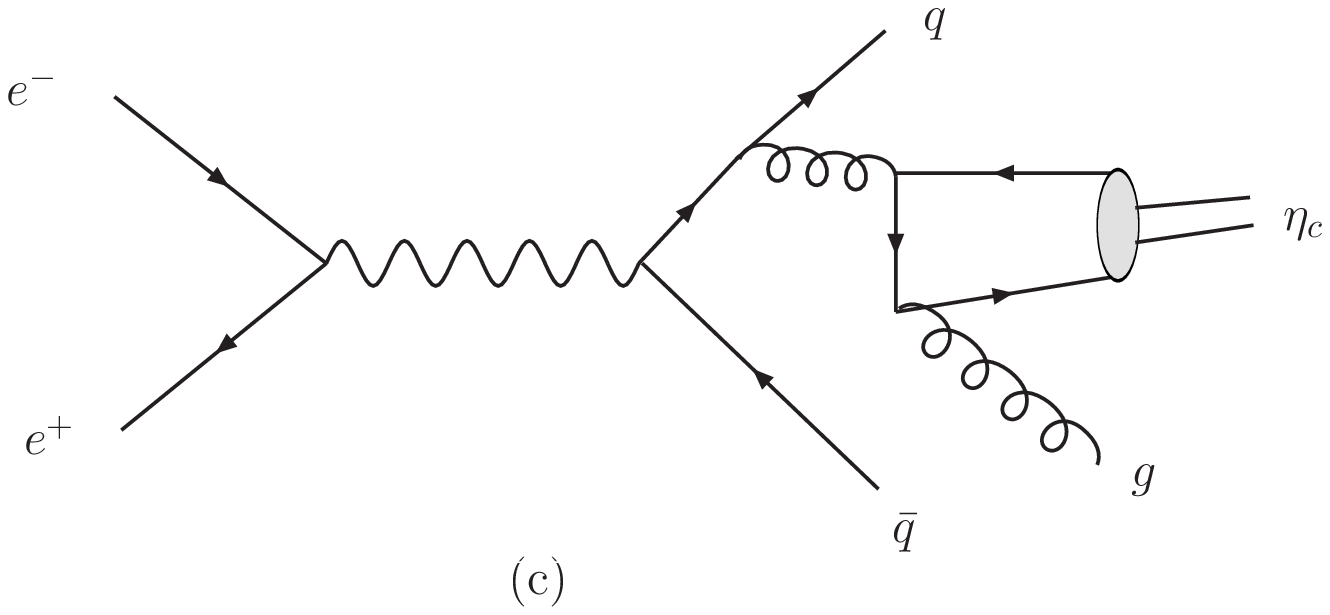   , width=5cm}
\caption{Representative diagrams for $\eta_c$ 
  electroproduction via color-singlet transition: at order $\alpha_S^2$ in association with a $c\bar c$ pair (a), and at order $\alpha_S^3$ with three gluons (b) and with a light quark pair (c).}
\label{fig:ee-etacdiagr}
}
\FIGURE[t]{
\epsfig{file=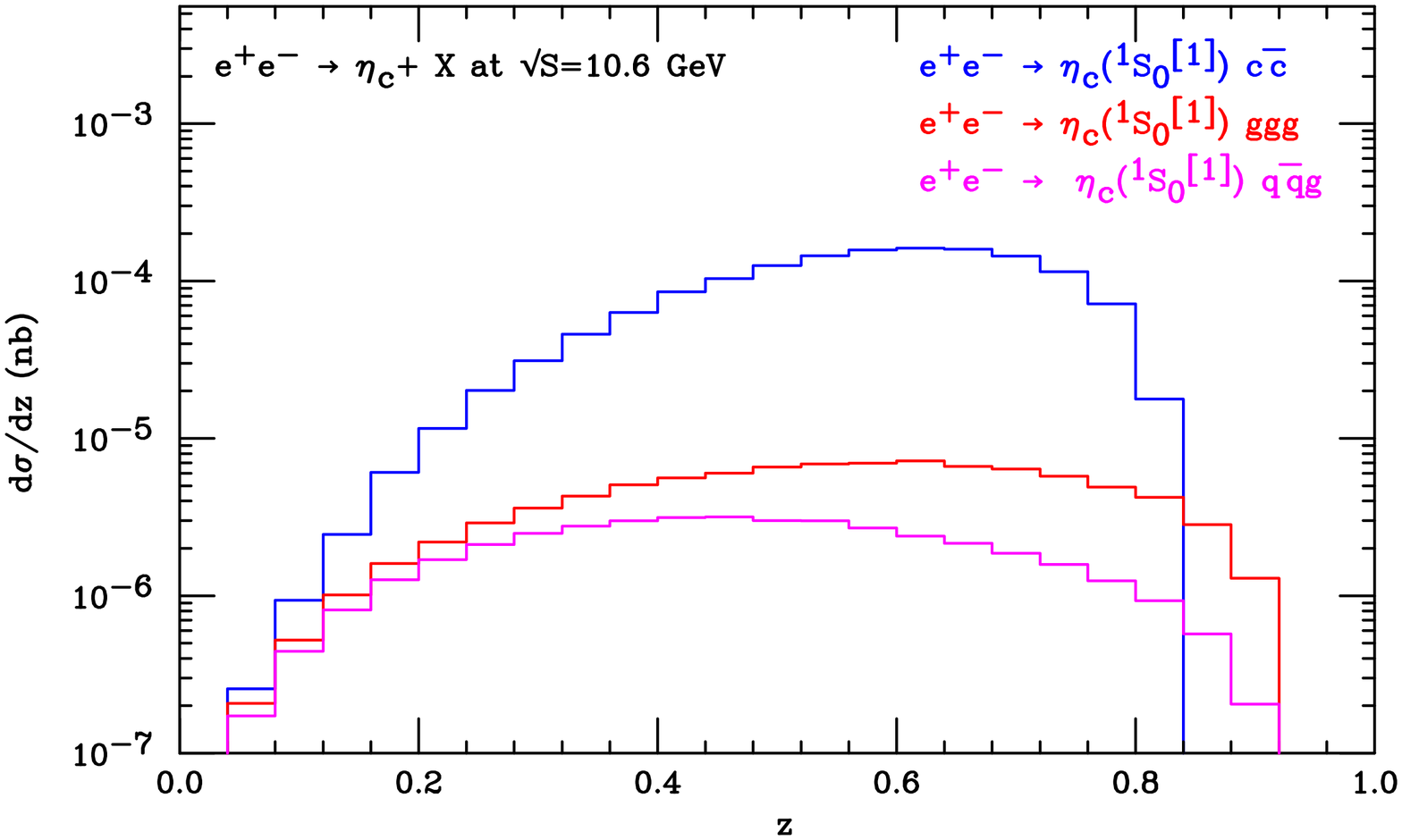, angle=90, width=12cm}
\caption{Momentum fraction of an $\eta_c$ in $e^+e^-$ collisions at $\sqrt{s}=10.6$ GeV.}
\label{fig:etacX}
}
\noindent As a result, the associated production is dominating the $\eta_c$ production
\begin{equation}
\frac{\sigma (e^+e^- \rightarrow \eta_c c \bar c)}{\sigma (e^+e^- \rightarrow \eta_c X)} = 91.6 \%
\end{equation}

It is interesting to note that the corresponding predictions for  
$J/\psi$ production are quite different. A similar calculation gives:
\begin{equation}
\sigma(e^+e^- \rightarrow J/\psi c\bar c)=148 \, \textrm{fb}, \qquad \sigma(e^+e^- \rightarrow J/\psi gg)=266 \, \textrm{fb}\,,
\end{equation}
and therefore, at leading order,
\begin{equation}
\frac{\sigma(e^+e^- \rightarrow J/\psi c\bar c)}{\sigma(e^+e^- \rightarrow J/\psi X)}=35.7 \%\,.
\end{equation}
In contrast to the above LO estimate, data indicate that
$J/\psi$ is produced in association with charm most of the time~\cite{Abe:2004ww}:
\begin{equation}
\frac{\sigma(e^+e^- \rightarrow J/\psi c\bar c)}{\sigma(e^+e^- \rightarrow J/\psi X)}|_{EXP}=82 \pm 0.15 \pm 0.14  \%\,.
\end{equation}
It would be interesting to have the analogous measurement for the $\eta_c$ too.

\subsection{$p\bar p \to \Upsilon + X $ at the Tevatron}


As a last application of our code, we consider the contributions
coming from higher order processes to the inclusive production 
of a $\Upsilon(^3S_1^{[1]})$ at the Tevatron. In particular, we focus on
the prediction for $p_T$ distribution.

It is easy to verify that the partonic differential cross section at
LO, Fig.~\ref{fig:upsilonCS}(a), falls off very quickly with the $p_T$, as $1/p_T^8$,
due to the presence of two very off-shell heavy quark propagators in all contributing diagrams.  At NLO $^3S_1^{[1]}$ production in association with gluons features 
diagrams  with only one off-shell heavy quark whose contribution scale as 
$1/p_T^6$, Fig.~\ref{fig:upsilonCS}(b).  At NNLO both
fragmentation contributions, Fig.~\ref{fig:upsilonCS}(c), and high-energy
enhanced contributions, Fig.~\ref{fig:upsilonCS}(d), appear.  While a full
NNLO calculation including virtual contributions is beyond our present technical
capabilities, one can argue that the terms with a different scaling,
such as $1/p_T^6$ and $1/p_T^4$ appear at the first time at tree-level
at order $\alpha_S^4$ and $\alpha_S^5$ respectively, and therefore
they are finite and are not affected by the (missing) virtual contributions
taking place at the same order.\footnote{This approach can be
explicitly validated for the $1/p_T^6$ terms using the full NLO
calculation~\cite{Campbell:2007ws}.  A detailed phenomenological analysis is
in progress~\cite{progress}.} 
\FIGURE[ht]{
\epsfig{file=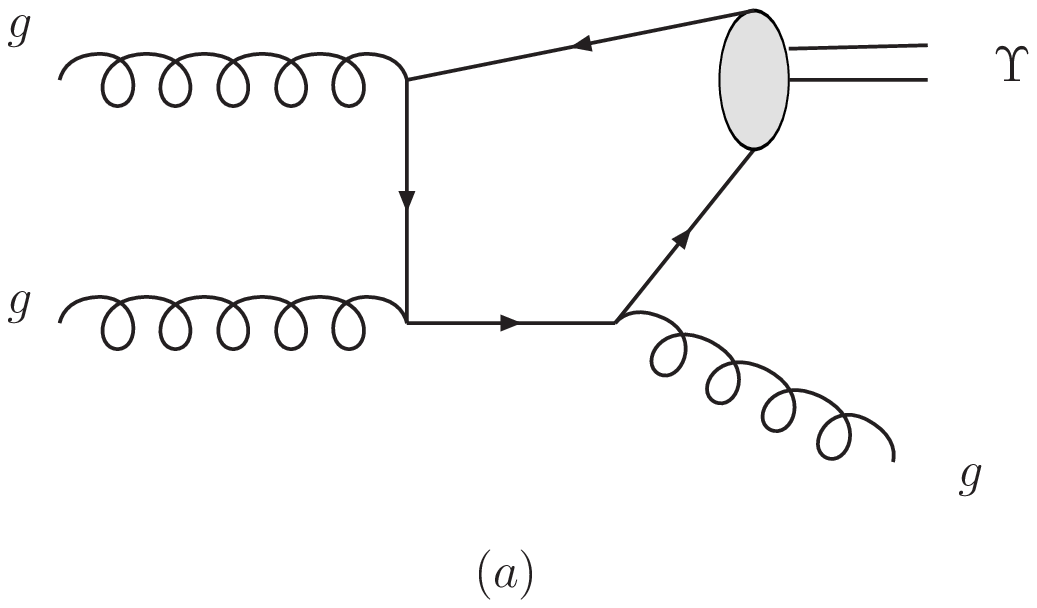 ,width=6cm}
\hspace*{1cm}
\epsfig{file=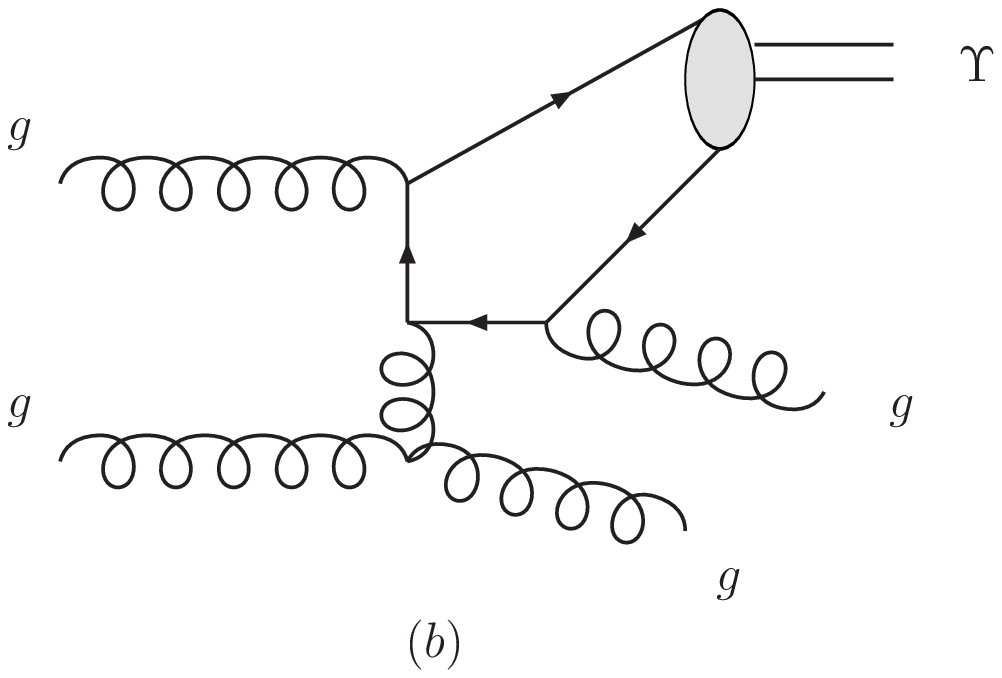   , width=6cm}
\hspace*{0.5cm}
\epsfig{file=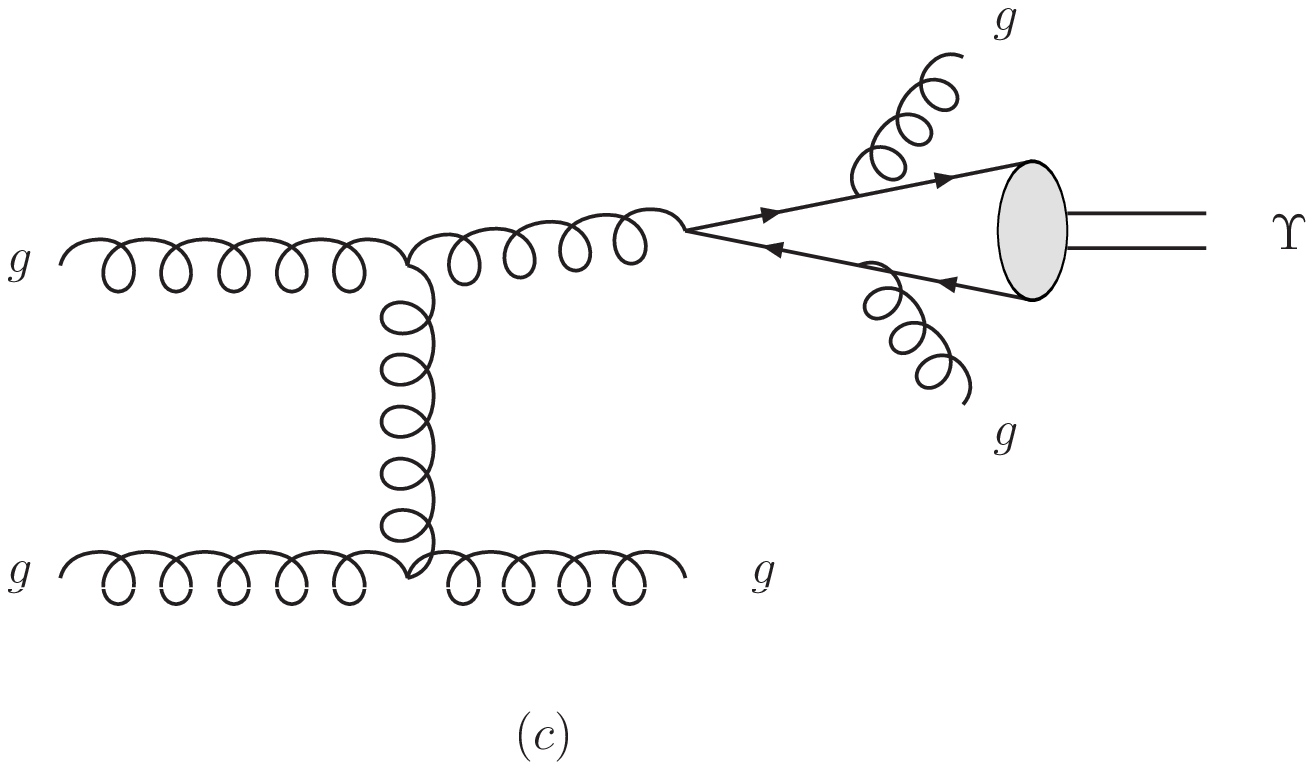   , width=7.0cm}
\hspace*{0.5cm}
\epsfig{file=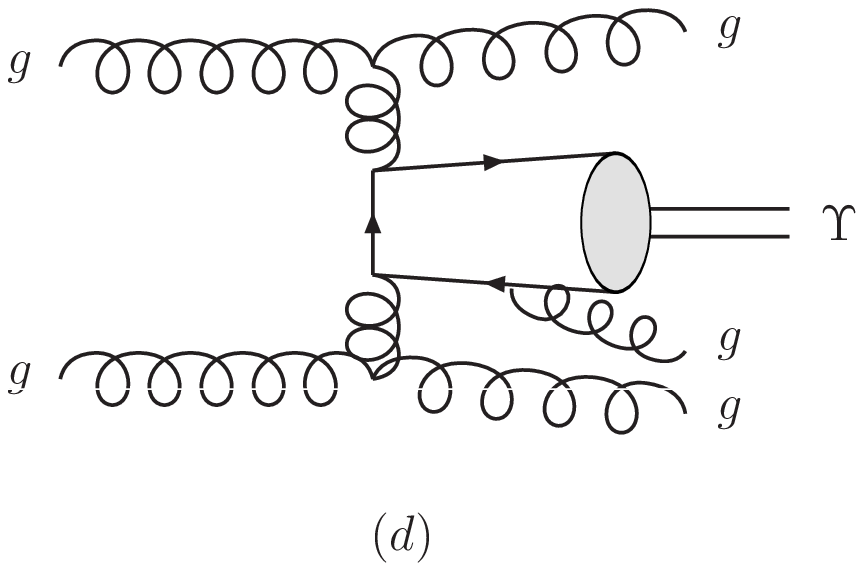   , width=5.8cm}
\caption{Tree-level contributions to $\Upsilon$ hadroproduction 
via a color-singlet transition (a) at LO $\alpha_S^3$, (b) at NLO $\alpha_S^4$, (c,d) at NNLO $\alpha_S^5$.}
\label{fig:upsilonCS}
}
Even at
tree-level, the calculation of multi-parton amplitudes is very
challenging and an automatized computational method is mandatory.  
For instance, the amplitude for $gg \to ~^3S_1^{[1]} + 3 g$
involves thousands of Feynman diagrams which reduce to several
hundreds when the color and spin projections are applied. Another
complication comes from the large number of different parton-parton
subprocesses, whose bookkeeping becomes quite cumbersome for large multiplicities. 
The above difficulties are easily solved by MadGraph. The identification of
the relevant $ parton\, parton  \to ~^3S_1^{[1]} + 3 \,partons$ subprocesses
and the generation of the corresponding amplitudes is matter of a few seconds. In
Fig.~\ref{fig:ptY} we show the different scaling in $p_T$ for the
various gluonic amplitudes $gg \to ~^3S_1^{[1]} + n g$, with
$n=1,2,3$. A minimal invariant mass for any gluon (initial or final
state) pairs, $s_{ij}>4m_b^2$, is required to avoid 
the phase space regions where the matrix element is singular. The $p_T$ shapes, however, depends weakly on the cutoff.

\section{Conclusions and Outlook}

\FIGURE[ht]{
\epsfig{file=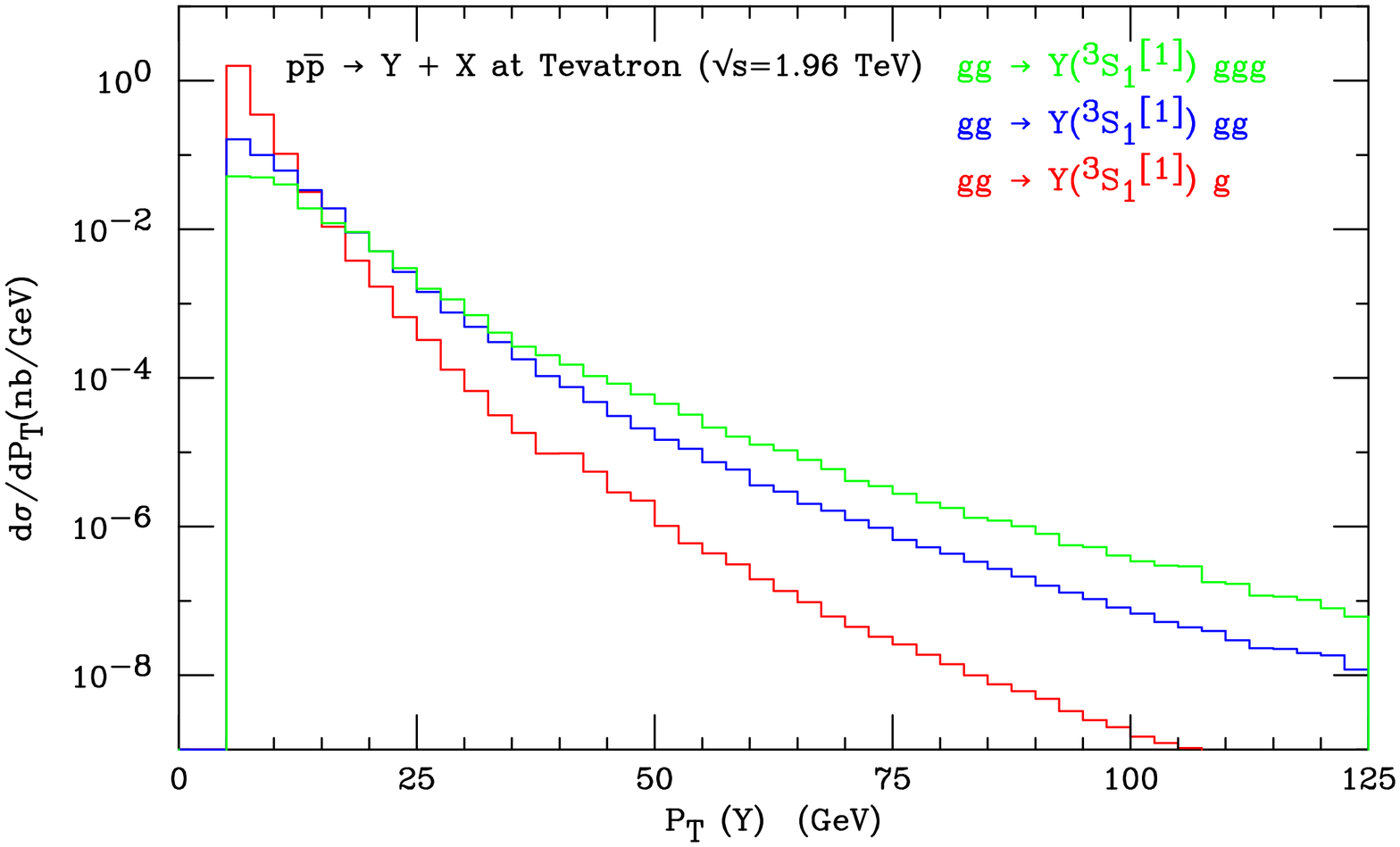 , angle=90, width=12cm}
\caption{Transverse momentum distribution for $\Upsilon$ 
production at the Tevatron, Run II. 
A cut on the minimal invariant mass of each pair of  initial and 
final state gluons, $s_{ij}>4 m_b^2$, is applied to avoid soft and 
collinear divergences. Only amplitudes involving gluons are included. 
Factorization and renormalization scales are fixed to 
$\mu_F=\mu_R=2 m_b$, with $m_b=4.75$ GeV, $\langle {\cal O}(Y)\rangle=9.28$ GeV$^3$.
The PDF set is CTEQ6M and $\alpha_S(2 m_b)=0.180$.}
\label{fig:ptY}
}

Quarkonium physics is a rich and active field of research, where our
understanding of QCD can be challenged and tested. Collider data, in
particular, suggests that our present description of the quarkonium
production mechanism is not fully satisfactory yet.  Given the
importance that $\Upsilon$ and $J/\psi$ production will have at the
LHC not only as calibration tools, but also for bottom and even top
physics, it is mandatory to improve the reliability  and the
flexibility of our predictions.  In this work we have completed the
first step towards the development of a multi-purpose Monte Carlo
generator for quarkonium physics, i.e., a code for the automatic
evaluation of any tree-level matrix element involving a heavy-quark
pair in a definite spin and color state.  To illustrate its
possibilities we have presented several applications to collider
physics phenomenology.

The following step will be to interface the matrix element generator
to MadEvent for the automatic phase space integration and event
generation. With such a code, events at the parton-level for any
process of interest can be generated and then passed through Pythia or
Herwig for the parton shower and hadronization and finally to detector
simulation. This will allow not only a much wider range of possible
studies, such as for example pattern of extra radiation in quarkonium
events, but also more direct flow of information from theory to
experiment in testing new ideas or new approaches (such as different
scaling rules for the various non-perturbative matrix elements). Work
in this direction is in progress.

\section*{Acknowledgments}

We are thankful to Geoff Bodwin and Eric Braaten for many useful discussions.
We would also like to thank all members of CP3 for the great atmosphere 
and enviroment that foster our efforts. This work is partially supported 
by the US National Science Foundation (Contract number NSF PHY 04-26272 ),
by the Belgian Federal Science Policy (IAP 6/11). P.A. is  Research Fellow 
of the Fonds National de la Recherche Scientifique, Belgium.

\bibliography{database}

\end{document}